\documentclass[reprint, amsmath,amssymb,aps]{revtex4-2}

\usepackage{float}
\usepackage{graphicx}
\usepackage{dcolumn}
\usepackage{bm}
\usepackage{xcolor}
\usepackage{cancel}
\usepackage{siunitx}
\DeclareSIUnit{\dBm}{dBm}

\begin{document}

\title{All-optical nonlinear activation function based on stimulated Brillouin scattering}
\author{Grigorii Slinkov$^{1, *}$, Steven Becker$^{1, 2, *}$, Dirk Englund$^{3}$, and Birgit Stiller$^{1, 2,\dagger}$\\
\small{ \textcolor{white}{blanc\\}
$^{1}$Max-Planck-Institute for the Science of Light, Staudtstr. 2, 91058 Erlangen, Germany\\
$^{2}$Department of Physics, Friedrich-Alexander-Universität Erlangen-Nürnberg, Staudtstr. 7, 91058 Erlangen, Germany\\
$^{3}$Research Laboratory of Electronics, Massachusetts Institute of Technology, Cambridge, Massachusetts 02139, USA\\
$^*$these authors contributed equally, $^\dagger$corresponding author: birgit.stiller@mpl.mpg.de}}
\date{\today}
\begin{abstract}
Photonic neural networks have demonstrated their potential over the past decades, but have not yet reached the full extent of their capabilities.
One reason for this lies in an essential component -- the nonlinear activation function, which ensures that the neural network can perform the required arbitrary nonlinear transformation.
The desired all-optical nonlinear activation function is difficult to realize, and as a result, most of the reported photonic neural networks rely on opto-electronic activation functions.
Usually, the sacrifices made are the unique advantages of photonics, such as resource-efficient coherent and frequency-multiplexed information encoding.
In addition, opto-electronic activation functions normally limit the photonic neural network depth by adding insertion losses.
Here, we experimentally demonstrate an in-fiber photonic nonlinear activation function based on stimulated Brillouin scattering.
Our design is coherent and frequency selective, making it suitable for multi-frequency neural networks.
The optoacoustic activation function can be tuned continuously and all-optically between a variety of activation functions such as \textsc{LeakyReLU}, \textsc{Sigmoid}, and \textsc{Quadratic}.
In addition, our design amplifies the input signal with gain as high as $\SI{20}{\dB}$, compensating for insertion losses on the fly, and thus paving the way for deep optical neural networks.
\end{abstract}
\maketitle
\section*{Introduction}
Artificial neural networks (ANNs) have emerged as powerful instruments for solving difficult tasks that range from speech recognition to image processing in medicine.
Thanks to their self-learning abilities and nonlinearity~\cite{abiodun_state---art_2018}, they can provide creative solutions derived from their training on large data sets.
After years of rapid scaling of model complexity, machine learning inference and training are close to reaching a bottleneck formed by the limitations of conventional Boolean logic processing hardware, especially with regard to power consumption, latency, and data movement.
\begin{figure*}[ht]
\centering
\includegraphics[width=.99\textwidth]{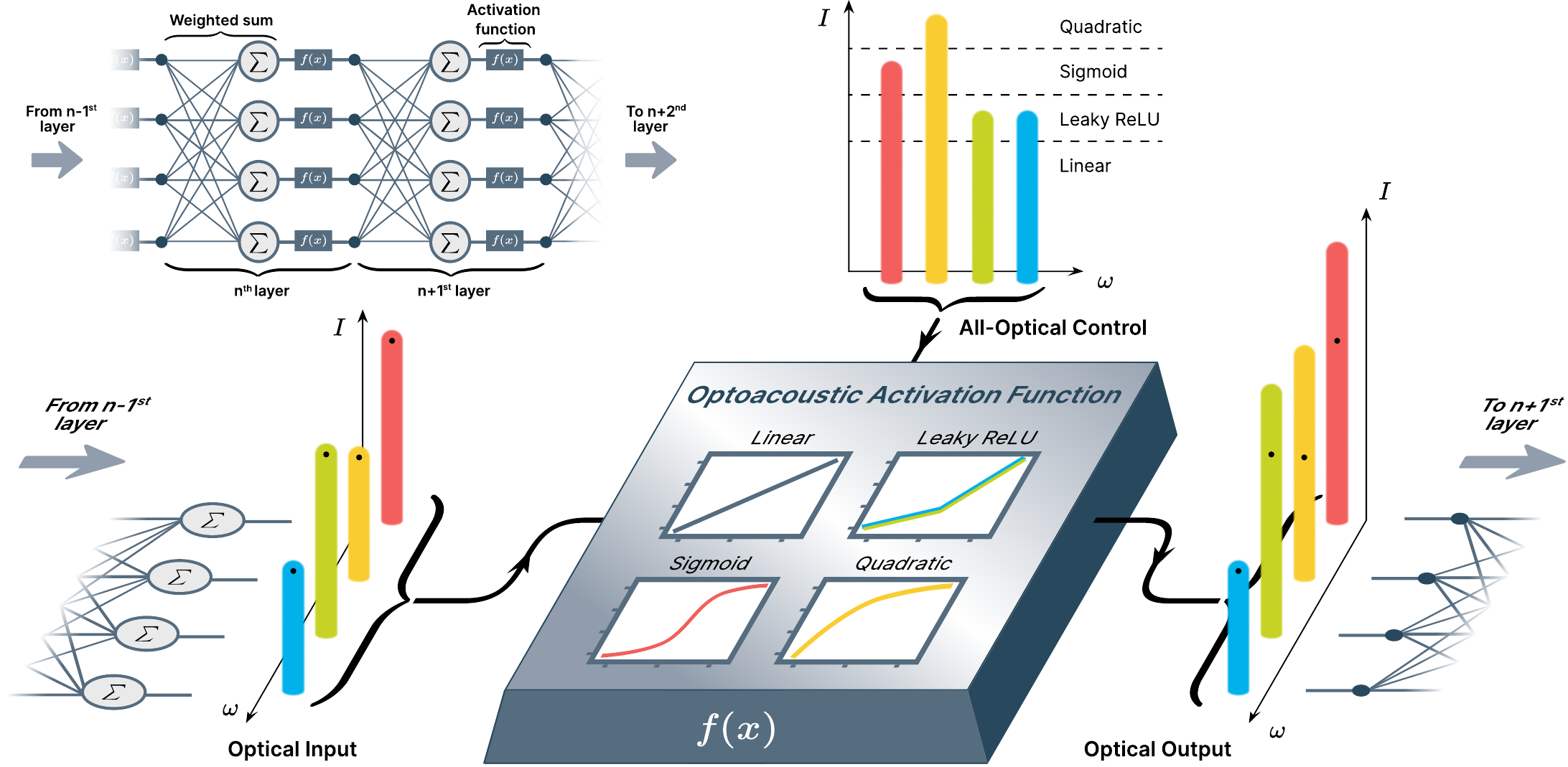}
\caption{A schematic representation of an optoacoustic activation function, employed between layers $n-1$ and $n+1$ of an all-optical multi-frequency neural network.
The result of the weighted summation of the $n-1^\mathrm{st}$ layer is sent to the optoacoustic activation function together with the multi-frequency control signal.
The magnitudes of the control signal's spectral components define the type of activation function applied to the corresponding input signal component.
The optical output of the optoacoustic activation function has the same frequency components as the input, with their magnitudes nonlinearly transformed depending on the type of activation function.
The output is fed to the next layer of the neural network.
The inset shows a conceptual scheme of a neural network. Each neuron performs two crucial operations: it takes a weighted sum of the inputs and applies a nonlinear activation function to the result.}
\label{fig:schematic_illustration}
\end{figure*}
Overcoming this ``von Neumann'' bottleneck has motivated the search for new ANN computing architectures based on fundamentally different principles.
Transferring the linear algebraic operation -- vector-matrix multiplication -- of ANNs to the optical domain, in particular, yields a potential for fundamental improvements in energy consumption and latency~\cite{sande_advances_2017, bogaerts_programmable_2020, wu_lithography-free_2023, valensise_large-scale_2022, sludds_delocalized_2022, shen_deep_2017, chen_deep_2023, bandyopadhyay_single_2022, davis_iii_frequency-encoded_2022, shastri_photonics_2021}.
Although these approaches have demonstrated the potential of optical neural networks (NNs), most of them achieve nonlinearity through opto-electro-optic conversion or digital post-processing~\cite{valensise_large-scale_2022, shen_deep_2017, chen_deep_2023,  bandyopadhyay_single_2022, davis_iii_frequency-encoded_2022, sludds_delocalized_2022, pour_fard_experimental_2020, jha_reconfigurable_2020, williamson_reprogrammable_2020}.
As the opto-electronic conversion of the signal at each neuron limits the power efficiency, the computing speed, and the scalability of the system~\cite{miscuglio_all-optical_2018},
all-optical activation functions are demanded~\cite{mourgias-alexandris_all-optical_2019, feldmann_all-optical_2019, shi_nonlinear_2022, miscuglio_all-optical_2018}.
The desirable, but so far not demonstrated in one ``package'', features of an all-optical activation function are:
\textbf{(i)} programmable nonlinearity, \textbf{(ii)} low insertion loss, \textbf{(iii)} coherence, \textbf{(iv)} frequency selectivity, and \textbf{(v)} compatibility with on-chip designs.
A programmable activation function allows the optical NN to adapt better to a specific problem and can be used as an additional training parameter.
It has been shown for digital NNs that this additional degree of freedom is beneficial for the NN's performance~\cite{agostinelli_learning_2015, dubey_activation_2022}.
Insertion losses limit the depth of the neural network, reducing the number of layers that can be stacked before the signal will have to be amplified.
A coherent all-optical activation function is not only beneficial
for phase-based optical NN architectures, such as~\cite{shen_deep_2017, bandyopadhyay_single_2022}, but can also allow to implement efficient training
schemes~\cite{lopez-pastor_self-learning_2023}.
So far, to the best of our knowledge, the only reported coherent activation function is~\cite{shi_nonlinear_2022}.
However, this architecture does not show features \textbf{(ii)} and \textbf{(iv)} because it attenuates the signal with a saturable-absorber-like response and cannot discriminate between different input frequencies.
The latter is an essential of applying resource-efficient frequency-basis information encoding, which is a unique feature of photonics, to optical NNs.
Due to the overall lack of a frequency-selective activation function, multi-frequency photonic machine learning architectures have been so far limited to vector-matrix multiplication~\cite{davis_iii_frequency-encoded_2022, feldmann_parallel_2021, buddhiraju_arbitrary_2021}.

Here we experimentally demonstrate an optoacoustic activation function that combines features \textbf{(i)}-\textbf{(v)} (see Fig.~\ref{fig:schematic_illustration}).
Our design is based on the nonlinear effect of stimulated Brillouin scattering (SBS)~\cite{wolff_brillouin_2021, boyd_nonlinear_2008, kobyakov_stimulated_2010}, which arises from the interplay between optical and acoustic fields.
Brillouin scattering can be easily observed in optical waveguides, such as optical fibers and on-chip devices~\cite{merklein_chip-integrated_2017, van_deventer_polarization_1994}.
SBS is inherently frequency-selective~\cite{stiller_cross_2019}, which makes it particularly suitable for resource-efficient frequency-basis information encoding.
This means that our all-optical activation function treats different frequencies independently, while being coherent.
Moreover, our approach amplifies the signal with a positive net gain, facilitating its use in deep optical NNs.
The nonlinear response is controlled all-optically and can be tuned continuously between different activation function shapes, including \textsc{LeakyReLU}, \textsc{Sigmoid} and \textsc{Quadratic},
which are favored by the machine learning community~\cite{dubey_activation_2022, valensise_large-scale_2022}.

While Brillouin scattering has been traditionally used for lasers~\cite{gundavarapu_sub-hertz_2019, otterstrom_silicon_2018, chauhan_visible_2021}, sensing~\cite{galindez-jamioy_brillouin_2012, geilen_extreme_2023}, gyroscopes~\cite{li_microresonator_2017, lai_earth_2020}, and microscopy~\cite{antonacci_recent_2020, prevedel_brillouin_2019, scarcelli_confocal_2008}, with our work we reveal new applications, going beyond what has been previously demonstrated for Brillouin-based signal processing~\cite{becker_optoacoustic_2023, zeng_nonreciprocal_2022, marpaung_integrated_2019,eggleton_brillouin_2019, zhu_stored_2007, merklein_chip-integrated_2017, becker_high-speed_2023}.
Our approach is not limited to a specific platform as SBS can be observed in different waveguide types ranging from optical chips~\cite{van_laer_interaction_2015, kittlaus_large_2016, shin_control_2015, merklein_-chip_2021, munk_surface_2019, gyger_observation_2020, botter_guided-acoustic_2022, ye_surface_2023, rodrigues_-chip_2023} and microresonators~\cite{bahl_observation_2012, kim_non-reciprocal_2015,cryer-jenkins_second-order_2023, enzian_non-gaussian_2021, enzian_observation_2019} to photonic crystal fiber~\cite{dainese_stimulated_2006, zeng_nonreciprocal_2022, pang_stable_2015} and fiber tapers~\cite{beugnot_brillouin_2014, xu_strong_2023}.

In the following, we demonstrate different nonlinear input-output mappings of the photonic activation function and its tunability.
In addition, we apply it to a dual-frequency signal, with a $\SI{3}{\giga\hertz}$ channel separation, demonstrating its frequency selectivity.

\section{Concept}
\begin{figure}[hbt]
\centering
\includegraphics[width=.5\textwidth]{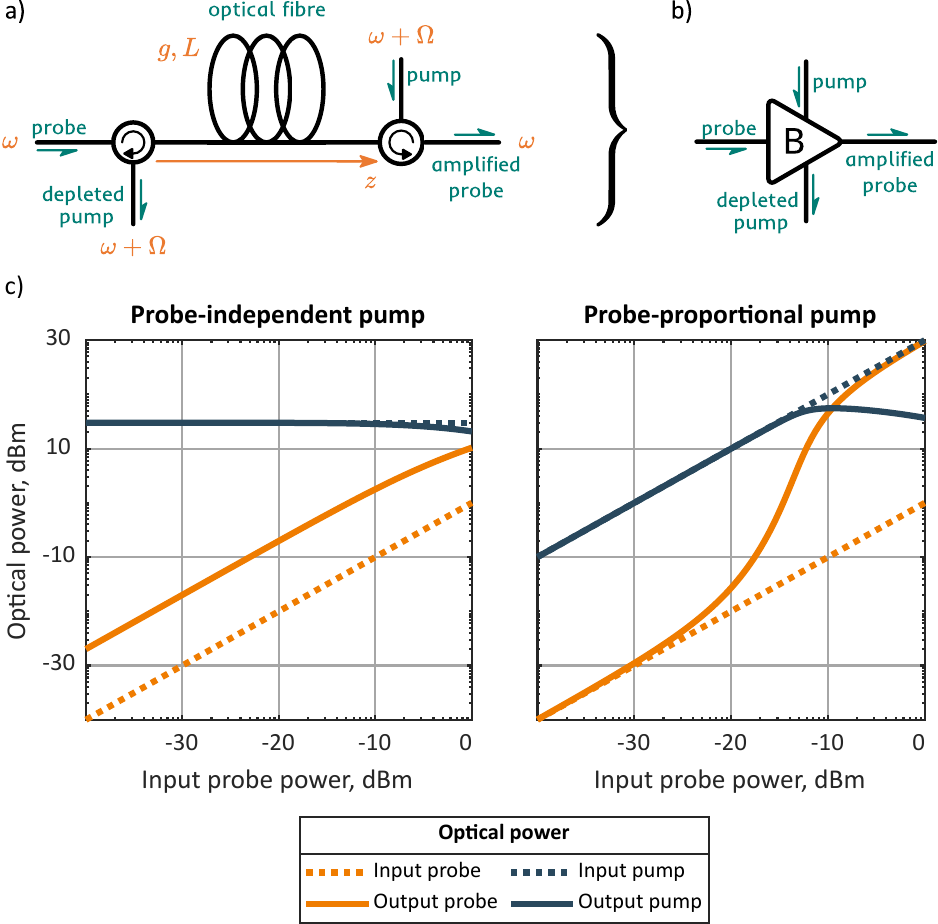}
\caption{\textbf{a)} A schematic of an optical fibre-based Brillouin amplifier. \textbf{b)} A convention introduced to represent the Brillouin amplifier. \textbf{c)} Numerical simulation of SBS process, showing linear (left panel) and nonlinear (right panel) relation between the input and the output probe optical power.}
\label{fig:pump-probe}
\end{figure}

Stimulated Brillouin scattering (SBS) is a third-order nonlinear effect that couples a pair of counterpropagating optical waves with a traveling acoustic wave serving as a mediator between them~\cite{wolff_brillouin_2021}. It follows a strict phase-matching condition~\cite{stiller_cross_2019}: the frequencies of the optical waves propagating in opposite directions have to be separated by the acoustic wave's frequency $\Omega$, which, for a given optical wavelength, is defined by the properties of the interaction medium. A schematic of an experimental realization is depicted in Figure~\ref{fig:pump-probe} \textbf{a}, where the probe wave $a_\mathrm{probe}$ is taken to be the one with the lower frequency $\omega$ and the pump wave  $a_\mathrm{pump}$  oscillates with the frequency $\omega + \Omega$. The interaction between the fields $a_\mathrm{probe}$, $a_\mathrm{pump}$, and $b$  can be described formally with the interaction Hamiltonian~\eqref{eq: recurrent_interaction_Hamilton}~\cite{zhang_quantum_2023}:
\begin{align}
    \label{eq: recurrent_interaction_Hamilton}
    \begin{aligned}
    \hat{H}_\mathrm{int} = \hbar g \, \int_{-\infty}^{\infty} \mathrm{d}z \, \left(\hat{a}_\mathrm{pump}\, \hat{a}^{\dagger}_\mathrm{probe} \, \hat{b}^{\dagger}  + \, \mathrm{H.c.}\right),
    \end{aligned}
\end{align}
with the optoacoustic coupling constant $g$ and the time- and space-dependent wave packet operators $\hat{a}_\mathrm{probe}(z,t)$, $\hat{a}_\mathrm{pump}(z,t)$, $\hat{b}(z,t)$ of the probe, pump and acoustic field, respectively. A detailed description of SBS can be found in the supplementary material.

Our nonlinear activation function is based on a modified Brillouin amplifier scheme that has been demonstrated to deliver high-gain and low-noise operation~\cite{pannell_stimulated_1993, xing_high-power_2008, pelusi_brillouin_2020}.
Under the undepleted pump assumption, the equations governing SBS allow for a simple analytic solution.
That is, when the pump is considered to be unaffected by the Brillouin interaction.
In this case, the relation between the optical powers of the probe $P_\mathrm{in}$, the pump $P_\mathrm{pump}$ and the amplified probe $P_\mathrm{out}$ can be written as follows~\cite{boyd_nonlinear_2008, gokhan_analytical_2018}:
\begin{equation}
    \label{eq: Brillouin_amp}
    P_\mathrm{out} = P_\mathrm{in} \, \exp\!\left(g_\mathrm{} L \cdot P_\mathrm{pump}  \right),
\end{equation}
where $g_\mathrm{}$ is the Brillouin gain and $L$ is the interaction length.
We consider now a slightly different situation, where the usually independent pump power is made dependent on the optical input of the Brillouin amplifier: $P_\mathrm{pump} = P_\mathrm{pump}(P_\mathrm{in}) = \gamma P_\mathrm{in}$.
Then, the Brillouin amplifier shows nonlinear input-output behaviour and equation~\eqref{eq: Brillouin_amp} would write as:
\begin{equation}
    \label{eq: Brillouin_amp_nonlin}
    P_\mathrm{out} = P_\mathrm{in} \, \exp\!\left(g_\mathrm{}L \cdot \gamma P_\mathrm{in} \right)
\end{equation}
Solving numerically the corresponding set of coupled mode equations for SBS allows us to study the behaviour of an input-dependent Brillouin amplifier for applying it as a versatile nonlinear activation function for an optical neural network architecture.
The corresponding equations can be found in the supplementary material.
The simulation results are presented in Fig.~\ref{fig:pump-probe} \textbf{c}. The $g\cdot L$ product is taken to be $\num{0.1}$ in our simulation. In both panels the input probe power $P_2(0)$ is swept from $\SI{-40}{}$ to $\SI{0}{\dBm}$.
In the left panel the input pump power $P_\mathrm{pump}$ is kept constant, which results in a linear dependence of probe output $P_\mathrm{out}$ on the probe input $P_\mathrm{in}$.
As the output probe power surpasses $\SI{-10}{\dBm}$, the pump starts to get depleted, and so the probe output growth rate decreases as well.
In the right panel both probe $P_\mathrm{in}$ and pump $P_\mathrm{pump}$ inputs are swept at the same rate.
As one can see, this changes drastically the output probe dynamic: the dependence goes from linear to exponential growth and then back to linear, as we begin to observe the pump depletion.
This nonlinear input-output relation in combination with the strict phase matching condition of SBS represents a perfect tool to implement a frequency-selective nonlinear activation function.

\section{Experimental results}
\subsection{Single-wavelength operation}
\begin{figure}[ht]
    \centering
    \includegraphics[width=.99\linewidth]{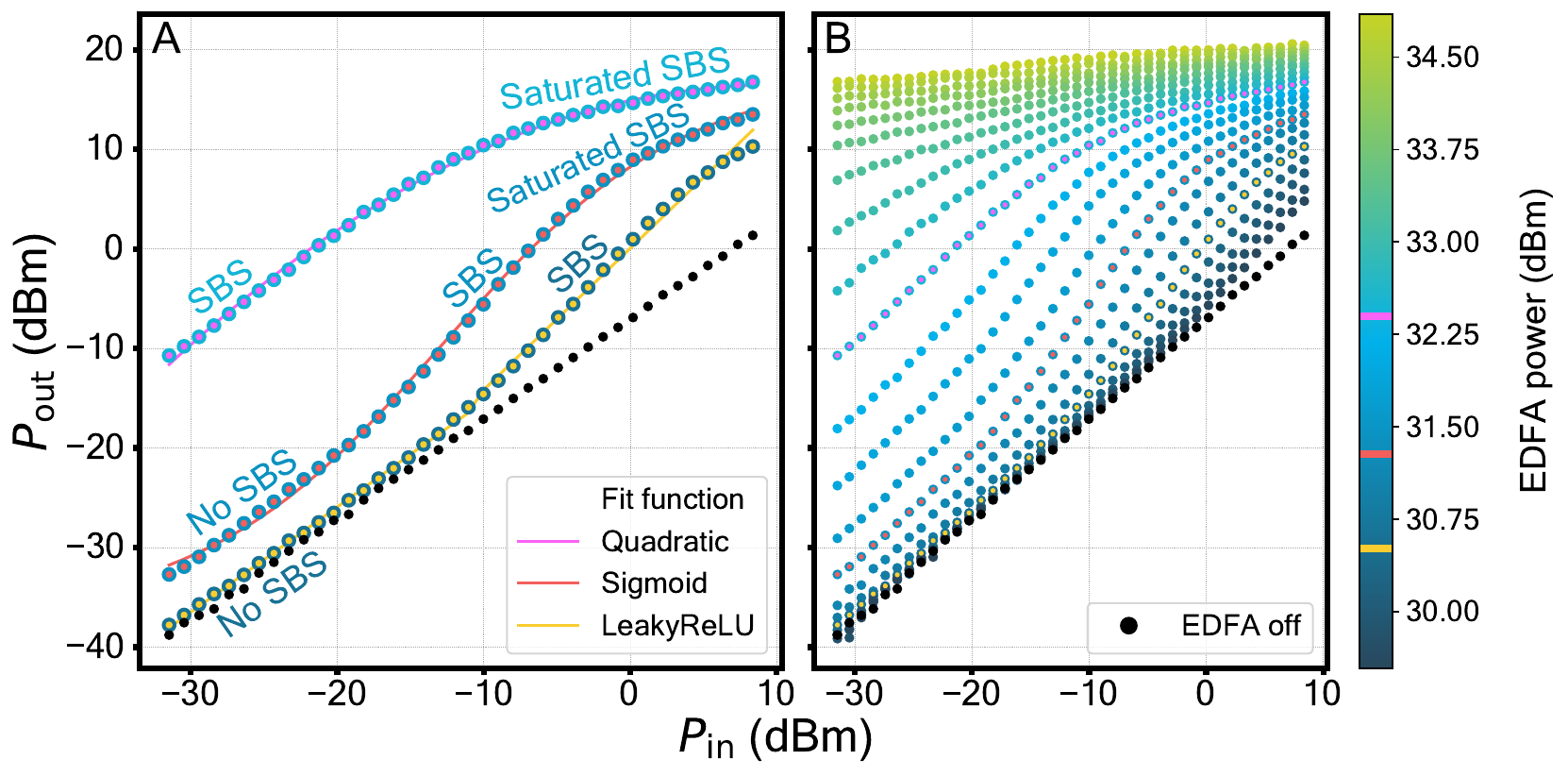}
    \caption{Nonlinear activation function shapes -- the mapping between the optical input and the optical output of the setup. \textbf{A} A selection of curves fitted with conventional analytic activation functions. \textbf{B} The complete family of curves obtained at various $1^\mathrm{st}$ stage pump optical power levels provided by the EDFA.}
    \label{fig: single_freq}
\end{figure}

The measurement results are presented in Fig.~\ref{fig: single_freq}, where the output power is plotted against the input power for different $1^{\mathrm{st}}$ stage pump power levels provided by an Erbium-doped fiber amplifier (EDFA).
Panel \textbf{A} shows a selection of activation function shapes accessible with the setup; panel B demonstrates the complete family of activation function curves. The variation of the EDFA power allows to choose a specific curve, adjusting the activation function shape continuously.
When the EDFA is turned off, the activation function is linear.

The selected curves in panel \textbf{A} are fitted with analytical activation functions. A pump power of \SI{30.5}{\dBm} corresponds to the \textsc{Leaky ReLU} function. Its first section is a direct proportionality between the input and the output in the absence of SBS process -- in this region the pump power is too low to be in the exponential regime of SBS.
When the corresponding pump power exceeds the Brillouin threshold, the
input light gets amplified. As a result, the input-output dynamic changes its slope as required for \textsc{Leaky ReLU}.
The next curve, obtained at pump power of \SI{31.3}{\dBm}, is fitted with \textsc{Sigmoid}. Its nonmonotonic growth can be split into three distinct sections. First, the absence of SBS in the beginning. Second, the amplification provided by SBS in the middle section of the plot. Lastly, the saturated SBS in the final section, where the SBS process becomes so intense that the pump starts to get depleted, resulting in the saturation of the growth. The last curve in Fig.\ref{fig: single_freq}~\textbf{A}, obtained at \SI{32.4}{\dBm} pump power, is fitted with a \textsc{Quadratic} function, formed by a SBS process that gives way to saturated SBS as the probe power is increased.
The optoacoustic activation function provides a gain of up to
$2.4\,\mathrm{dB}$, $8.8\,\mathrm{dB}$, and $21.9\,\mathrm{dB}$ for the
\textsc{Leaky ReLU}, \textsc{Sigmoid}, and \textsc{Quadratic} case, respectively. Hence, the optoacoustic activation function can be used to compensate for losses induced by the preceding matrix operation. This is an essential feature for implementing deep optical NNs.

\subsection{Dual-wavelength operation} \label{sec: dual}
We demonstrate the feature of frequency selectivity by splitting the input into two wavelength-multiplexed channels, provided by two tunable lasers at the input. The frequency separation between the two is set to \SI{3}{\giga\hertz}, limited in the experiment by the spectrum analyzing device resolution.

\begin{figure*}[ht]
\centering
\includegraphics[width=.99\linewidth]{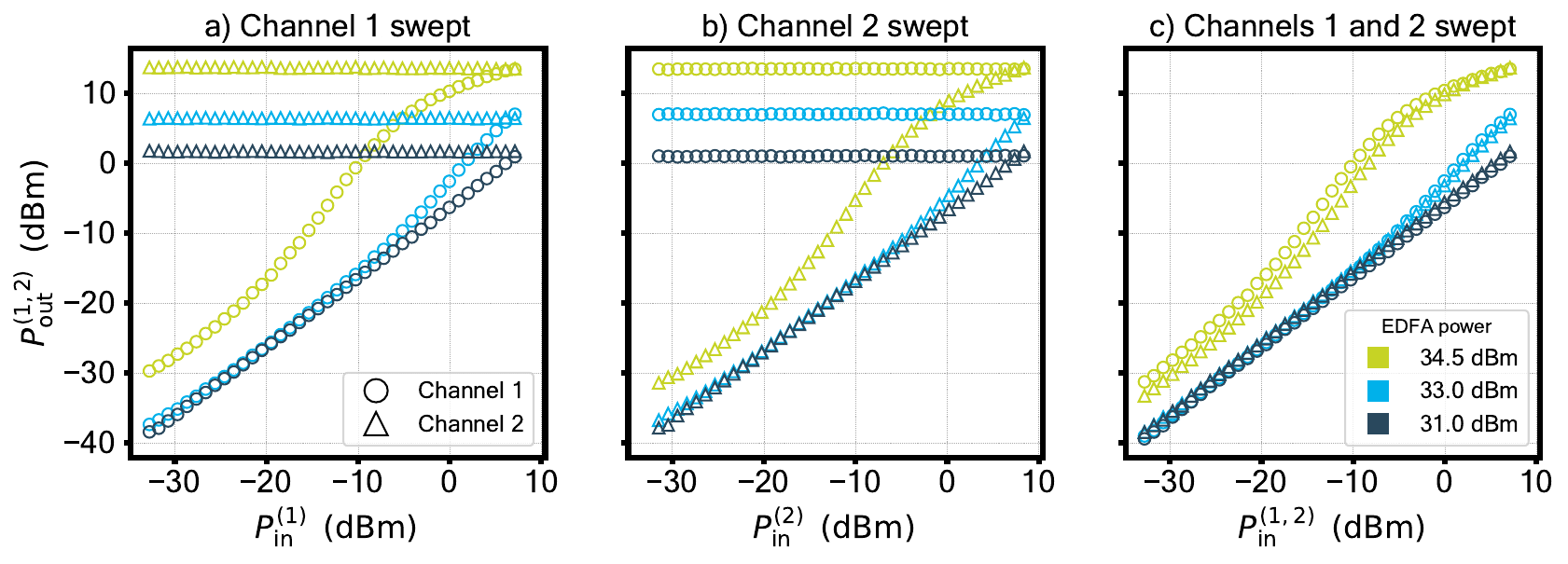}
\caption{Demonstration of the frequency-selective operation. Circles and triangles correspond to a pair of wavelength-multiplexed channels, provided by two lasers at the input. The shape of the activation function in either of channels is not affected by the SBS interaction taking place in the neighbouring channel.}
\label{fig: dual_freq}
\end{figure*}

Each of the two channels hosts its own variable attenuator, which allows us to control the power levels independently (see supplement for details). We use a Finisar WaveAnalyzer to perform a wavelength-selective measurement of optical power at the output. The measurement results are plotted in Fig.~\ref{fig: dual_freq}. Panel \textbf{a)} shows the case where the power in channel 1 is swept, while the power in channel 2 is kept constant, panel \textbf{b)} is vice versa. In panel \textbf{c)} both channels are swept simultaneously. The reference case where both channels are swept, but the EDFA is turned off can be found in the Supplement. The presented selection shows that for a given channel neither the presence, nor the variation of a signal in the neighbouring channel affects the shape of the nonlinear activation function.
The \textsc{Sigmoid} activation function shape achieved at maximum EDFA power in the dual-wavelength mode does not match the \textsc{Quadratic} shape achieved at maximum EDFA power in the single-wavelength mode.
This is due to EDFA distributing its output power equally between the pumps of the two frequency channels, yielding a \SI{3}{\dB} less power per channel.

\section{Discussion and conclusion}
We have experimentally demonstrated for the first time, to the best of our knowledge, a nonlinear photonic activation function based on stimulated Brillouin scattering.
Our activation function is coherent and frequency selective owing to the nature of SBS.
A coherent activation function is the next step from the phase-reliant optical matrix multiplication approaches \cite{bandyopadhyay_single_2022, shen_deep_2017}
to fully all-optical ANNs that would not require opto-electronic conversion, \textcolor{black}{which imposes bandwidth limitations, introduces cross-talk, requires introduction of time delay and eliminates frequency selectivity}.

The frequency selectivity opens up prospects of drastically increasing the data throughput by exploiting wavelength multiplexing techniques to, for example, distribute the neurons in a layer across the frequency domain.
It is the first demonstration of a frequency sensitive activation function, which shows no correlation between frequency channels that hinders existing solutions, such as frequency-encoded deep neural networks~\cite{davis_iii_frequency-encoded_2022}.
Moreover, our activation function could transform the existing multi-frequency photonic machine learning architectures (which have been so far limited to linear operations~\cite{feldmann_parallel_2021, buddhiraju_arbitrary_2021}) into actual multi-frequency ANNs.
As it has been shown, there is no cross-talk between neighbouring frequency channels at frequency separations as small as $\SI{3}{\giga\hertz}$,
which surpasses the telecommunication standard of $\SI{25}{\giga\hertz}$.
For the continuous wave case, the minimal frequency separation between the two channels is intrinsically limited by the linewidth of the optoacoustic gain function,
which, for the commercial single-mode fiber at room temperature is about $\SI{26}{\mega\hertz}$~\cite{le_floch_study_2003}.
As shown in~\cite{le_floch_study_2003, becker_high-speed_2023}, the linewidth can be decreased by lowering the temperature of the waveguide.
In the pulsed case, the minimal frequency separation is dictated by the pulse length~\cite{stiller_cross_2019}.
The general rule of thumb for SBS-based applications is this: channel separation has to be higher than
$\Delta \nu_\mathrm{laser}+1/\tau_\mathrm{pulse}+\Delta \nu_\mathrm{B}$, where $\Delta \nu_\mathrm{laser}$ is the laser linewidth, $\tau_\mathrm{pulse}$ is the pulse length and $\Delta \nu_\mathrm{B}$ is the acoustic gain linewidth.

The activation function can be tuned by varying the $1^\mathrm{st}$ Brillouin amplifier's pump power to take such well-proven shapes
as \textsc{LeakyReLU}, \textsc{Sigmoid}, and \textsc{Quadratic}.
As the dynamics of the $1^\mathrm{st}$ Brillouin amplifier can be controlled externally, it opens the possibility to use the nonlinearity as an additional training parameter of an optical ANN.
For digital ANNs, this has been shown to be a powerful tool for boosting the ANN performance~\cite{agostinelli_learning_2015, dubey_activation_2022}.
It is also feasible to engineer the $1^\mathrm{st}$ stage pump in such a way that different frequency channels would have different activation function shapes.

The output signal amplification that is inherent to our activation function design is suitable for compensating insertion and propagation losses.
This should be particularly useful for designing deep optical NNs that comprise multiple neuron layers~\cite{lopez-pastor_self-learning_2023}.

Though the experimental realisation presented in the paper relies on highly nonlinear optical fiber as the optoacoustic interaction medium,
our approach is not limited to this platform. Conventional single-mode fiber (SMF) and photonic crystal core fiber (PCF) are
also an eligible choice for SBS as well as integrated waveguides, including on-chip devices~\cite{eggleton_brillouin_2019, merklein_chip-integrated_2017, choudhary_advanced_2017, martinez_optoacoustic_2023, morrison_compact_2017}.
The choice of the platform combined with pulsed operation constitute the way for improving the energy efficiency of the presented optoacoutic activation function.
One needs to maximize the term $g\cdot P_\mathrm{pump}L_\mathrm{eff}-\alpha L$ in order to improve the
energy efficiency. Here, $P_\mathrm{pump}$ is the pump power, $L_\mathrm{eff}$ is the effective interaction length,
and $\alpha$ is the optical loss of the waveguide (see Suppl. for details).

In conclusion, our frequency selective and coherent photonic nonlinear activation fills a gap in the current landscape of photonic machine learning.
It could therefore be the key to unlocking the full potential of photonic neuromorphic computing.

\section{Methods}
\begin{figure}[tbh]
\centering
\includegraphics[width=.99\linewidth]{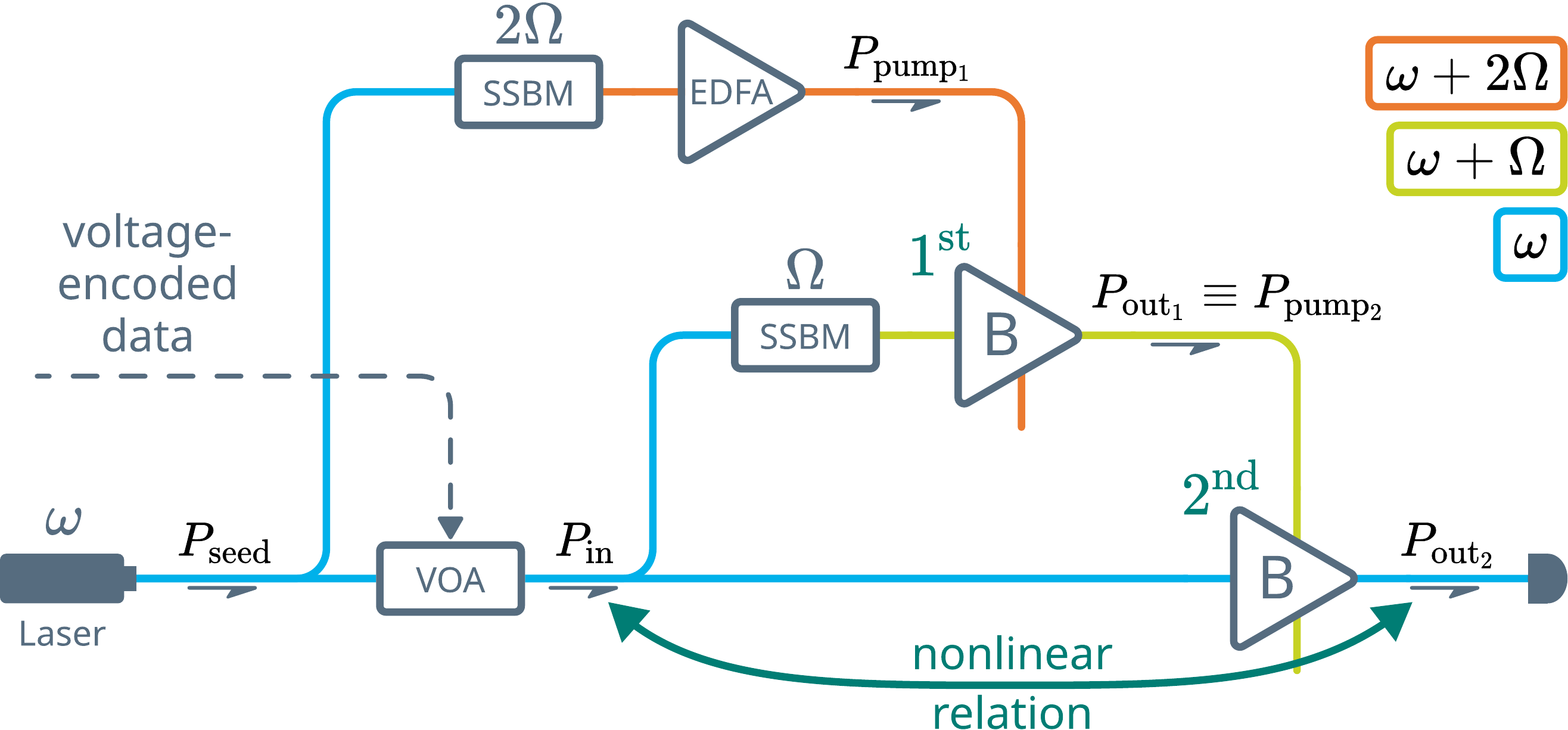}
\caption{A principal scheme of the experimental setup: VOA - voltage-operated attenuator, SSBM - single-sideband modulator, EDFA - Erbium-doped fiber amplifier, B - Brillouin amplifier, as introduced in Fig.~\ref{fig:pump-probe}~\textbf{b}. The color of the connecting lines depicts the light frequency.}
\label{fig:exp_scheme}
\end{figure}

We implement the photonic nonlinear activation using a setup depicted schematically in Fig.~\ref{fig:exp_scheme}.
We build what can be called a double-stage Brillouin amplifier: the output of one Brillouin amplifier ($1^{\mathrm{st}}$ stage) is utilized as a pump for another ($2^{\mathrm{nd}}$ stage).
The $1^{\mathrm{st}}$ stage Brillouin amplifier is pumped with an Erbium-doped fiber amplifier (EDFA).

The reason for pumping the $2^\mathrm{nd}$ stage with a Brillouin amplifier, as opposed to applying an EDFA directly, is that
conventional EDFAs operate in the saturated regime, providing input-independent output power.
This way, replacing the $1^\mathrm{st}$ stage with an EDFA would have eliminated any possible relation between the input and the pump that is required by \eqref{eq: Brillouin_amp_nonlin}.

The setup is fed with a $\SI{1550.12}{\nano\meter}$ fiber-coupled laser.
A voltage-controlled attenuator (VOA) is used in the experiment to test the nonlinear input-output behaviour of the setup, simulating the amplitude-encoded data from the previous neuron layer.
The VOA is inserted before the light gets distributed between the two stages, which ensures that they receive the same amplitude variation, as required by \eqref{eq: Brillouin_amp_nonlin}.
Note that feeding the top branch of the setup with the same laser is a measure taken to enhance the stability of the Brillouin amplifier and is not an actual requirement.

A Brillouin frequency shift $f_\mathrm{B}$ (corresponding angular frequency $\Omega=2\pi f_\mathrm{B}$) is applied to the middle branch of the setup, satisfying the SBS phase matching condition for the $2^{\mathrm{nd}}$ stage.
This requires the signal in the top branch to be up-shifted by the sum of the Brillouin frequencies of the fibers.
The optical fibre used in both of the stages was of the same material and structure, yielding a $2\Omega$ shift for the top branch.

We use highly nonlinear fiber (HNLF) with equal parameters for the two Brillouin amplifiers.
The lengths of the HNLF fibers used for the first and the second stage are $\SI{20}{\meter}$ and $\SI{100}{\meter}$, correspondingly.
The Brillouin frequency of the fibers is $f_\mathrm{B} = \SI{9.730}{\giga\hertz}$.
A couple of single sideband modulators (SSBMs) driven with two separate RF sources apply required frequency shifts to the top and the middle branches of the setup.
\newpage
\bibliographystyle{unsrt}
\bibliography{main.bib}

\end{document}


\title{Supplementary information: All-optical nonlinear activation function based on stimulated Brillouin scattering}
\author{Grigorii Slinkov$^{1, *}$, Steven Becker$^{1, 2, *}$, Dirk Englund$^{3}$, and Birgit Stiller$^{1, 2,\dagger}$\\
\small{ \textcolor{white}{blanc\\}
$^{1}$Max-Planck-Institute for the Science of Light, Staudtstr. 2, 91058 Erlangen, Germany\\
$^{2}$Department of Physics, Friedrich-Alexander-Universität Erlangen-Nürnberg, Staudtstr. 7, 91058 Erlangen, Germany\\
$^{3}$Research Laboratory of Electronics, Massachusetts Institute of Technology, Cambridge, Massachusetts 02139, USA\\
$^*$these authors contributed equally, $^\dagger$corresponding author: birgit.stiller@mpl.mpg.de}}
\date{\today}
\begin{abstract}
This document provides supplementary information to ``All-optical nonlinear activation function based on stimulated Brillouin scattering''. Detailed description of the stimulated Brillouin scattering effect, extended description of the experimental setup, details on energy efficiency and latency of the nonlinear activation function are presented.
\end{abstract}
\maketitle
\section{Detailed description of Brillouin scattering}
\begin{figure}[ht]
\centering
\includegraphics[width=1\linewidth]{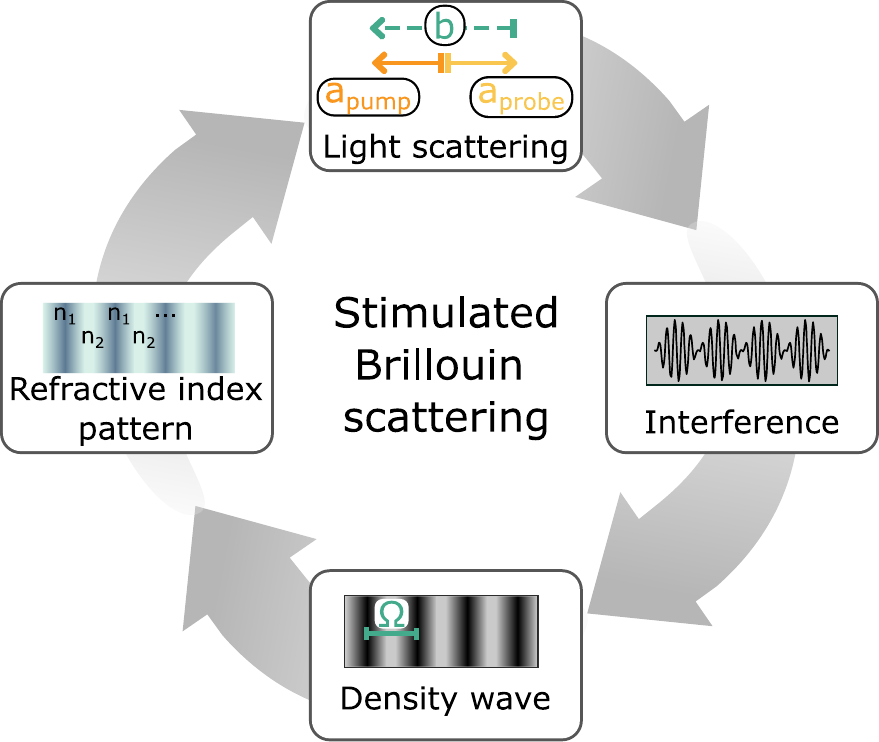}
\caption{Illustrated of the SBS feedback mechanism Initially, light waves $\omega_\mathrm{probe}$ and $a_\mathrm{pump}$ scatter off acoustic phonons $b$, experiencing a frequency shift known as Brillouin frequency $\Omega$. The interplay between the incident and backscattered light fields generates a dynamic interference pattern in motion. Electrosriction then translates this pattern into a mobile density wave, changing the refractive index $n$ of the medium due to the photoelastic effect. This resulting density wave augments scattering efficiency, enabling a stimulated process.}
\label{fig: illustration_SBS_feedback}
\end{figure}

In the following, we describe the underlying dynamic of stimulated Brillouin scattering (SBS) using Figure~\ref{fig: illustration_SBS_feedback}.
A more detailed description can be found in References~\cite{wolff_brillouin_2021, boyd_nonlinear_2008, kobyakov_stimulated_2010, agrawal_stimulated_2013}.
The coincidence of the two counter-propagating optical fields $a_\mathrm{probe}$ at frequency
$\omega_\mathrm{probe}$ and $a_\mathrm{pump}$ at frequency $\omega_\mathrm{pump}=\omega_\mathrm{probe} + \Omega$
in the waveguide creates an interference pattern. Through the effect of electrostriction, the interference pattern excites a traveling acoustic wave $b$.   
As the density variation is connected to the optical refractive index $n$ of the medium, the optical waves are thus exposed to a moving Bragg-grating-like refractive index pattern. 
This causes an inelastic scattering of the $a_\mathrm{pump}$ into the probe wave $a_\mathrm{probe}$ and the acoustic wave $b$ which then enhances the interference, substantiating the stimulated nature of the process.

\section{Extended illustration of the experimental setup} \label{sec: setup}
\begin{figure*}[ht]
\centering
\includegraphics[width=\textwidth]{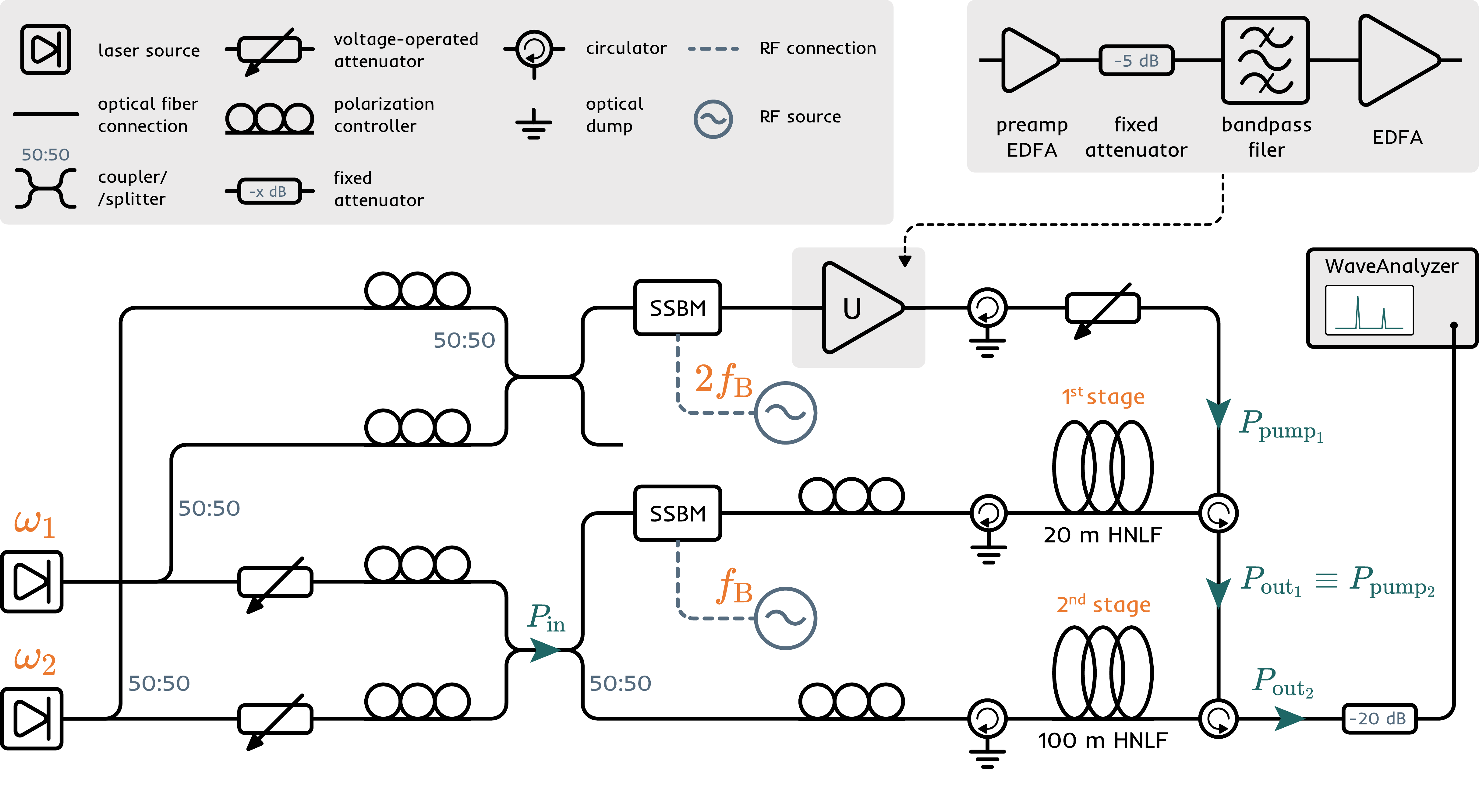}
\caption{Extended illustration of the experimental setup used in the study. In the single-frequency case one of the lasers was turned off.}
\label{fig: extended_experimental_setup}
\end{figure*}
The extended experimental setup used for both the single and dual-frequency study is shown in Figure~\ref{fig: extended_experimental_setup}. 
The setup is fed with a pair of lasers to simulate the frequency-multiplexed input. 
The separation between the two frequency channels $\omega_1-\omega_2\approx\SI{3}{\giga\hertz}$ is set by tuning the individual wavelengths of the lasers.
For the single-wavelength study the setup was the same, but only one of the two lasers was used.

Each laser's output is split using a 50:50 coupler. One half is then used for the nonlinear activation function, whereas the other serves as a seed for the optical pump branch. 
The input power of the nonlinear activation function is controlled by a voltage-operated attenuator (VOA). 
In the dual-frequency experiment a second VOA is used for each of the two lasers, ensuring independent control of optical power for each of the frequency channels.
The two frequency channels are combined using a 50:50 coupler. 
Their combination forms the activation function input, which will undergo the desired nonlinear transformation. 
The coupler acts also as splitter, feeding its input to the bottom and the middle branches of the setup which constitute the nonlinear activation function. 

The (dual-frequency) signal in the lower branch serves as the input for the $2^\mathrm{nd}$ stage Brillouin amplifier based on a $\SI{100}{\meter}$ long highly nonlinear fiber (HNLF).  
 
The (dual-frequency) signal in the middle branch is first up-shifted by the Brillouin frequency of the HNLF $f_\mathrm{B}= \SI{9.730}{\giga\hertz}$ (corresponding angular frequency $\Omega=2\pi f_\mathrm{B}$) with an optical IQ-modulator, which is configured to operate as a single-sideband modulator (SSBM).
The resulting frequency up-shifted signal gets amplified by the $1^\mathrm{st}$ stage Brillouin amplifier based on a $\SI{20}{\meter}$ long HNLF. 
After that, it serves as the pump for the $2^\mathrm{nd}$ stage Brillouin amplifier.

The top branch of the setup provides the optical pump for the  $1^\mathrm{st}$ stage Brillouin amplifier. Thus, it is seeded with the original laser combination, up-shifted by twice the Brillouin frequency and amplified by two cascaded Erbium-doped fiber amplifiers (EDFA). 
The first one acts as a pre-amplifier to seed the second high-power EDFA. 
Between the two EDFAs we apply a fixed value attenuator to comply with the input power limitations of the high-power EDFA. It is followed by a $\SI{1}{\nano\meter}$ bandpass filter to reduce the amplified spontaneous noise (ASE) level. 
A circulator isolates the EDFA from possible light backreflection and could be replaced with an optical isolator. 
The output power of the top branch is controlled with a third VOA for precise and quick tuning.

The output of the second Brillouin amplifier ($2^\mathrm{nd}$ stage) forms the output of the nonlinear activation function. It is detected with a WaveAnalyzer~\cite{noauthor_waveanalyzer_nodate} that performs a frequency-selective measurement. 

\section{Dual channel input-output reference}
\begin{figure}[h]
\centering
\includegraphics[width=\linewidth]{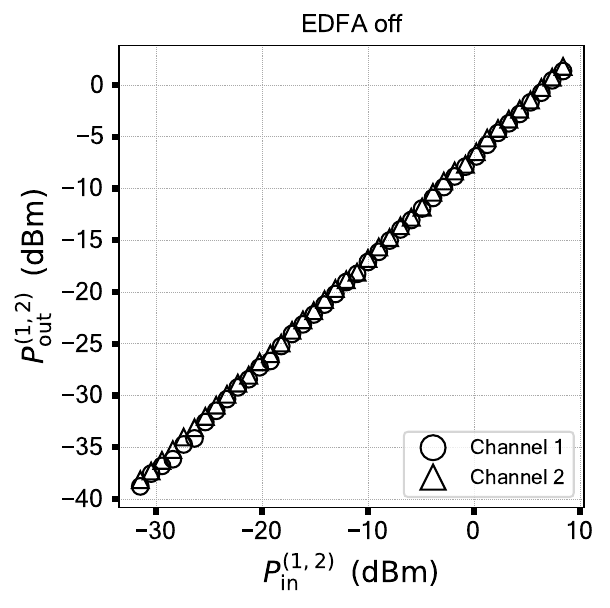}
\caption{Dual-frequency operation with the EDFA turned off.}
\label{fig: dual_EDFA_off}
\end{figure}
Figure \ref{fig: dual_EDFA_off} shows linear input-output dynamic for both frequency-multiplexed channels when the EDFA is turned off. It provides a reference for the Figure 4 of the manuscript. Power in both channels is swept.

\section{Numerical study details}
\begin{figure}[bh]
\centering
\includegraphics[width=0.6\linewidth]{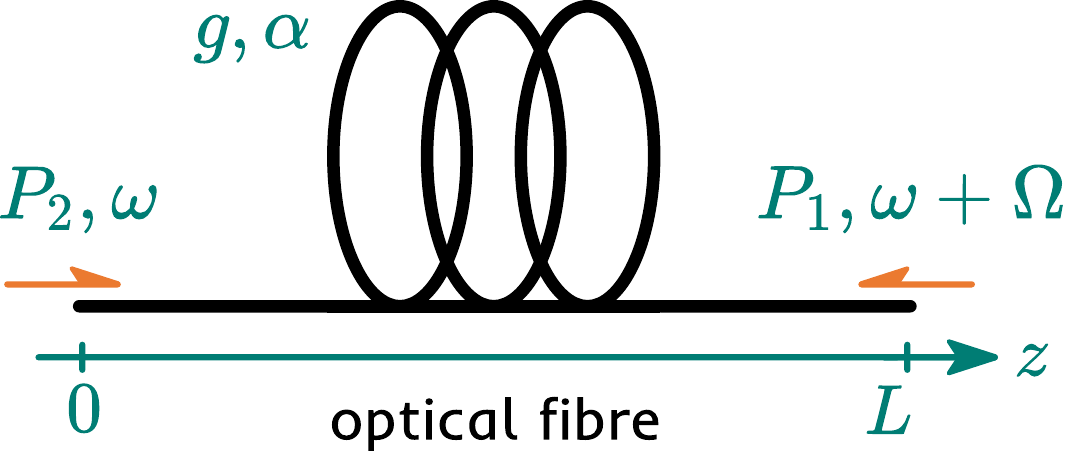}
\caption{Brillouin process in an optical fiber. $P_2$ is the probe wave, $P_1$ is the pump wave.}
\label{fig: B_fiber}
\end{figure}
In a unidimensional waveguide (optical fibre) the equations governing SBS process can be written in terms of pump and probe optical power.
Although these equations allow for an analytical solution~\cite{gokhan_analytical_2018}, we follow the numerical approach discussed in~\cite{boyd_nonlinear_2008}. 
We solve the set of equations~\eqref{diff_eqs} in the conservative limit (no absorption):
\begin{align}
\label{diff_eqs}
    \begin{cases}
        \frac{\partial P_1}{\partial z} &= -g P_1 P_2 \\
        \frac{\partial P_2}{\partial z} &= -g P_1 P_2,
    \end{cases}
\end{align}
where $P_1=P_1(z)$ and $P_2=P_2(z)$ are the pump and the probe power measured along the fiber length $z$ (Fig. \ref{fig: B_fiber}), correspondingly, which means that following the conventions introduced in Section~I of the main text,
\begin{align}
\label{boundary_cond}
    \begin{cases}
        P_1(L)=P_\mathrm{pump}\\
        P_2(0)=P_\mathrm{in}
    \end{cases}
\end{align}
Fig.~2 \textbf{c} of the main text presents the results of solving \eqref{diff_eqs} with boundary conditions \eqref{boundary_cond}.

\section{Energy-efficiency}
In order to reduce the power required to pump the first Brillouin amplifier and
therefore enhance the overall energy efficiency of the activation function,
one can choose a longer waveguide with higher optoacoustic gain $g$.

Firstly, this will reduce the minimum pump power required to transfer from the spontaneous to the stimulated regime which can be approximately with 
Equation~\eqref{eq: power_threshold_SBS}~\cite{agrawal_stimulated_2013}.
\begin{equation}
\label{eq: power_threshold_SBS}
P_\mathrm{th} = \frac{21 K}{g L_\mathrm{eff}},
\end{equation}
with the effective interaction length $L_\mathrm{eff}=\left(1-\mathrm{exp}\left({-\alpha L}\right)\right)/\alpha$, the optical loss $\alpha$, the effect of copolarized light $K=3/2$~\cite{van_deventer_polarization_1994}, and the peak Brillouin gain $g$.
In our case, we can assume $\alpha_\mathrm{HNLF} = \SI{0.28}{\per\kilo\meter}$, $g_\mathrm{HNLF}=\SI{1.25}{\per\watt\per\meter}$~\cite{becker_high-speed_2023}, which yields $P_\mathrm{th} = \SI{1.26}{\watt}$ and $P_\mathrm{th} = \SI{256}{\milli\watt}$ for the first and the second Brillouin amplifier, respectively.

Secondly, a higher gain length product will reduce the amount of pump power to achieve similar gain values as demonstrated.
In the simplest case, the gain of a Brillouin amplifier can be written as in Equation~\eqref{eq: SBS_gain_with_loss}.
\begin{equation}
    \label{eq: SBS_gain_with_loss}
    g_\mathrm{amp} = \exp\!\left(g L_\mathrm{eff} P_\mathrm{pump}- \alpha L\right)
\end{equation}
with the intrinsic Brillouin gain $g$ and the waveguide length $L$. Accordingly, the pump power $P_\mathrm{pump}$ can be reduced by increasing the Brillouin gain $g$ and $L_\mathrm{eff}$, which could be achieved with chalcogenide waveguides such as rib chips and fibers or photonic crystal fibers~\cite{eggleton_brillouin_2019, martinez_optoacoustic_2023}.

The current HNLF-based design with the first Brillouin amplifier length $L=\SI{20}{\meter}$ offers $10\,\mathrm{log}_{10}\left(g_\mathrm{ amp}\right)\approx 54\,\mathrm{dB}$ gain driven with $P_\mathrm{pump} = \SI{500}{\milli\watt}$.
In a chalcogenide waveguide, given the experimentally demonstrated  parameters $g_\mathrm{AsS}=\SI{500}{\per\watt\per\meter}$, $\alpha_\mathrm{AsS}=\SI{1.2}{\per\kilo\meter}$ and chip length of $L_\mathrm{AsS}=\SI{0.2}{\meter}$~\cite{eggleton_brillouin_2019,choudhary_advanced_2017},
the same gain can be achieved with $80\,\%$ less pump power required: $P_\mathrm{pump, AsS} \approx \SI{0.12}{\watt}$.
Note that even chip designs with $g_\mathrm{AsS}=\SI{750}{\per\watt\per\meter}$ have been shown experimentally, however with a shorter waveguide length~\cite{morrison_compact_2017}.
\section{Latency of the optoacoustic activation function}
The latency of the optoacoustic nonlinear activation function is dictated by the time of flight $\tau$ of the the $2^\mathrm{nd}$ stage Brillouin amplifier. 
Assuming that this device has the length $L_\mathrm{NLA}=L_\mathrm{waveguide} + L_\mathrm{components}$ which is written as the sum of the legnth of the SBS-active waveguide $L_\mathrm{waveguide}$, such as the HNLF, and the length of the components $L_\mathrm{components}$ which feed the light into the waveguide and out - for instance, the circulators. 
Then the ToF can be written as Equation~\eqref{eq: time_of_flight}.
\begin{align}
\label{eq: time_of_flight}
\tau_\mathrm{NLA}=L_\mathrm{NLA} n_\mathrm{eff} / \mathrm{c_0}
\end{align}
with the effective refractive index of the waveguide $n_\mathrm{eff}$ and the speed of light $\mathrm{c_0}$.
For the nonlinear activation function used in our study assuming $L_\mathrm{NLA, HNLF}\approx \SI{108}{\meter}$ and $n_\mathrm{eff}=1.44$ yields a latency of $\tau_\mathrm{NLA, HNLF} \approx \SI{500}{\nano\second}$.
\newpage
\bibliographystyle{unsrt}
\bibliography{sm.bib}